\documentclass[letter]{aa} 
\usepackage[colorlinks=true,linkcolor=blue,citecolor    = blue]{hyperref}
\usepackage{graphicx}
\usepackage{txfonts}
\usepackage{float}
\usepackage{fancyhdr}
\usepackage{multirow}
\usepackage{bigstrut}
\usepackage[flushleft]{threeparttable}
\usepackage{amsmath}
\usepackage{enumitem}
\usepackage{amssymb}
\usepackage[utf8]{inputenc}
\usepackage{etoolbox}
\usepackage{appendix}
\usepackage{url}
\usepackage{hypcap}
\usepackage{dirtytalk}
\usepackage{fontawesome}

\usepackage{booktabs}
\usepackage[graphicx]{realboxes}
\usepackage{adjustbox}
\usepackage{rotating,tabularx}
\usepackage{afterpage}
\usepackage{tablefootnote}
\usepackage{placeins} 
\usepackage{tikz}

\usepackage{bigstrut}

\definecolor{lime}{HTML}{A6CE39}
\DeclareRobustCommand{\orcidicon}{%
    \begin{tikzpicture}
    \draw[lime, fill=lime] (0,0) 
    circle [radius=0.16] 
    node[white] {{\fontfamily{qag}\selectfont \tiny ID}};
    \draw[white, fill=white] (-0.0625,0.095) 
    circle [radius=0.007];
    \end{tikzpicture}
    \hspace{-2mm}
}

\foreach \x in {A, ..., Z}{%
    \expandafter\xdef\csname orcid\x\endcsname{\noexpand\href{https://orcid.org/\csname orcidauthor\x\endcsname}{\noexpand\orcidicon}}
}
\newcommand{\orcid}[1]{\href{https://orcid.org/#1}{\textcolor[HTML]{A6CE39}{\orcidicon}}}
\usepackage{graphicx}

\newcommand{\gaia}{\textit{Gaia}}

\newcommand{\MIStetson}{$-3.948 ^{+0.037}_{-0.034}$\,mag}
\newcommand{\MIStetsonRelativeError}{$1.6\%$}
\newcommand{\MRp}{$-3.807 ^{+0.041}_{-0.035}$\,mag}
\newcommand{\MRpRelativeError}{$1.8\%$}

\newcommand{\muLMCRp}{$18.447^{+0.036}_{-0.042}$\,mag} 
\newcommand{\relMuLMCRp}{$1.8\%$}
\newcommand{\muSMCRp}{$18.898^{+0.049}_{-0.054}$\,mag} 
\newcommand{\relMuSMCRp}{$2.4\%$}

\setcitestyle{notesep={: }}

\begin{document}

\title{Calibrating the Tip of the Red Giant Branch and measuring Magellanic Cloud distances to $2\%$ exclusively with {\it Gaia}}
\titlerunning{{\it Gaia} DR3 TRGB calibration and LMC distance}

   \author{
   Mauricio Cruz Reyes \inst{1}\orcid{0000-0003-2443-173X}          \and Richard I. Anderson\inst{1,2}\orcid{0000-0001-8089-4419}    \and Bastian Lengen\inst{1}\orcid{0009-0007-8211-8262} } 
  
   \institute{Institute of Physics, \'Ecole Polytechnique F\'ed\'erale de Lausanne (EPFL), Observatoire de Sauverny, 1290 Versoix, Switzerland \\
    \email{mauricio.cruzreyes@alumni.epfl.ch, richard.anderson@epfl.ch}   }

\authorrunning{Cruz Reyes, Anderson \& Lengen}

   \date{Received 13 February 2026}

\abstract
{We have calibrated the Tip of the Red Giant Branch (TRGB) using our recent catalog of homogeneous, high-accuracy Globular Cluster (GC) distances. The GC distances were determined by a global joint fit to optical period-Wesenheit relations of their member RR Lyrae stars and type-II Cepheids, anchored by trigonometric parallaxes; all data taken from the ESA \gaia\ mission's (early) third data release (GDR3). Using $I-$band measurements in 48 GCs from P.~Stetson's database, we determined $M_{I,0}=$\MIStetson\ (\MIStetsonRelativeError\ in distance). Calibrating the TRGB using \gaia's homogeneous, space-based $RP$ photometry of $53$ GCs, we found $M_{RP,0}=$\MRp\ (\MRpRelativeError). The stated uncertainties include statistical and systematic effects, including the correlated nature of the GC distances. The robustness of our calibrations is demonstrated via tests against small-number statistics and analysis choices. Specifically, we found no significant metallicity effect for our sample of old, low-metallicity GCs. We measured $\sim 2\%$ distances to the Large (LMC) and Small Magellanic Clouds (SMC), \muLMCRp\ ($48.9\pm 0.9$\,kpc) and \muSMCRp\ ($60.2\pm 1.4$\,kpc), respectively, using a single well calibrated photometric system: $RP$ (spectro-)photometry from GDR3. Our new TRGB distances, whose absolute scale derives from \gaia\ parallaxes, are fully independent of the well-known detached eclipsing binary (DEB) distances and agree with them to within the uncertainties. Combining our new TRGB and existing DEB distances, we illustrate how additional constraints may be incorporated in the Local Distance Network and obtain $H_0=73.52\pm0.80\, \mathrm{km\,s^{-1}\,Mpc^{-1}}$.
Expected improvements due to the upcoming fourth \gaia\ data release are discussed.}
  

   \maketitle

\section{Introduction}
The Tip of the Red Giant Branch (TRGB) is an empirical feature in the color-absolute magnitude diagrams (CaMDs) of old stellar populations, and it is well understood to originate from the Helium flash in first-ascent red giant stars \citep{2002PASP..114..375S}. The TRGB provides a useful standard candle for distance determination \citep{1993ApJ...417..553L} and for measuring the Hubble constant, $H_0$, \citep{2024arXiv240317048L}. The TRGB is the most commonly used extragalactic stellar distance indicator \citep{2021AJ....162...80A}, and it is calibrated using known distances, preferably ones determined using geometrical methods. Since the TRGB magnitude is measured as the inflection point of the RGB luminosity function (LF) rather than using individual stars, this calibration most frequently relies on stellar populations observed at a common distance \citep[for an exception, see][]{2023ApJ...950...83L}. In particular, the detached eclipsing binary (DEB) distances to the Large (LMC) and Small Magellanic Clouds (SMC) \citep{2019Natur.567..200P,2020ApJ...904...13G} and the water megamaser distance to NGC\,4258 \citep{2019ApJ...886L..27R} have thus far enabled the most accurate TRGB calibrations. Galactic Globular Clusters (GCs) have also been frequently used to calibrate the TRGB, notably since the publication of trigonometric parallaxes from the (early) third data release of the ESA \gaia\ mission \citep[henceforth: GDR3]{GaiaMission,GEDR3,GDR3}. In particular, the catalog of literature GC distances compiled in \citet{2021MNRAS.505.5957B} has been frequently adopted for this purpose, even though the reported distances average correlated measurements (e.g., multiple distances based on the same set of RR Lyrae stars) alongside standard mean errors. Here, we seek to improve upon the Galactic TRGB calibration by anchoring it to our recent catalog of 93 GC distances measured from a joint fit to period-Wesenheit relations of a large number of RR Lyrae stars and type-II Cepheids anchored to GDR3 parallaxes of GCs \citep{2025arXiv250916331L,2024A&A...684A.173C}. 

The \citet{2025arXiv251023823H} recently presented a $1\%$ $H_0$ measurement based on the Local Distance Network (LDN), a novel and extensible system of equations that combines high-quality low-redshift distance constraints in a statistically rigorous manner. The LDN's absolute scale is anchored by geometrically measured distances, including the LMC's DEBs (optionally also the SMC's), the NGC4258 megamaser, and \gaia\ parallaxes of classical Cepheids as well as average parallaxes of open clusters hosting Cepheids. Creating additional linkages to the absolute scale provided by \gaia\ parallaxes is crucial for further improvements to the LDN's robustness and $H_0$ precision. The second goal of the present work was therefore to connect the LDN with the absolute scale of the GC distances from \citet{2025arXiv250916331L} by measuring the distances to the LMC and SMC using TRGB measurements performed in a single well calibrated photometric system unaffected by atmospheric effects. \gaia's $RP-$band is very well suited to this end, given its effective wavelength similar to $I-$band, and since it provides very precise photometry near the TRGB both in Galactic GCs and the Magellanic Clouds \citep{2024ApJ...963L..43A,2024ApJ...974..181K}. These new  distances may, in turn, serve as independent cross-checks of the DEB distances and provide an example for how to incorporate new constraints into the LDN.

\section{Data and Methodology}\label{sec:data}
We adopted GC distances from our recent global fit \citep{2025arXiv250916331L} to optical period-Wesenheit relations of $802$ RRab and $345$ RRc stars, as well as $21$ type-II Cepheids residing in 93 GCs, which relied on mean magnitudes from \citep{GDR3RRL,GDR3Cep}. This global solution is anchored by accurate average parallaxes of 37 GCs from \citet{2024A&A...684A.173C} determined using trigonometric parallaxes from GDR3 in the optimal magnitude range where residual parallax bias \citep{2021A&A...649A...4L} is negligible and taking into account the error floor due to angular covariance. 

Metallicities, [Fe/H], and color excesses, $E(B-V)$, of GCs were taken from \citet{2010arXiv1012.3224H}. Extinctions towards GCs were calculated assuming $A_I = 2.213\,E(B-V)$, $A_{RP} = 2.237 \, E(B-V)$ for $R_V = 3.3$. In the Magellanic Clouds, we adopted extinction-corrected apparent TRGB magnitudes in GDR3 $RP-$band from \citet{2024ApJ...974..181K}, which used color excesses, $E(V-I)$, from the \citet{2021ApJS..252...23S} red clump star reddening maps and $A_{RP}=1.486\,E(V-I)$  in the LMC (assuming $R_V=3.3$) and $A_{RP}=1.320\,E(V-I)$  in the SMC (assuming $R_V=2.7$). All extinction coefficients were computed for a typical star near the TRGB using \texttt{pysynphot} assuming the \citet{1999PASP..111...63F} reddening law  following \citet{2022A&A...658A.148A}.

TRGB magnitudes were determined in three photometric systems: GDR3 $RP$, Cousins $I-$band determined by photometric transformations of GDR3 data (cf. App.\,\ref{fig:photometric_transfromation_bright}), and ground-based $I-$band observations. To this end, we compiled GDR3 photometry of GC member stars \citep{2021MNRAS.505.5978V} in the GDR3 $RP$, $BP$, and $G-$bands \citep{2021A&A...649A...3R}, as well as in Cousins $I-$band by cross-match ($2"$ search radius) with the September 2025 version of the \citet{2019MNRAS.485.3042S} database\footnote{\url{https://www.canfar.net/storage/list/STETSON/homogeneous/Latest_photometry_for_targets_with_at_least_BVI}}. Since the TRGB is populated by the brightest objects in a GC, we followed P. Stetson's advice and carefully considered possible issues due to non-linearity or saturation, cf. Appendix~\ref{sec:photometry}. However, no significant problems were identified.

We constructed a joint de-reddened CaMD using the samples of stars and GCs that passed data selection criteria explained in App.\,\ref{app:data_selection} and listed in Tab.\,\ref{tab:selection_criteria}. From the CaMDs, we constructed smoothed LFs using a GLOESS kernel of width $\sigma_s=0.125$\,mag, sampled in bins of $0.004$\,mag, cf. Appendix \ref{sec:smoothing} for robustness tests against a range of GLOESS smoothing parameters. We measured absolute TRGB magnitudes from the resulting LFs as the inflection point of an unweighted [-1,0,1] Sobel filter following \citet{2024ApJ...963L..43A}, ensuring consistency with the LMC and SMC measurements and avoiding bias due to differing TRGB contrast. Uncertainties were determined by bootstrap resampling taking into account correlations among the GC distances. 
Appendix~\ref{sec:convergence} presents further detail and convergence tests.

\section{A $1.6\%$ TRGB calibration based on Galactic GCs}\label{sec:results}
\begin{figure}
    \centering
    \includegraphics[width=\linewidth]{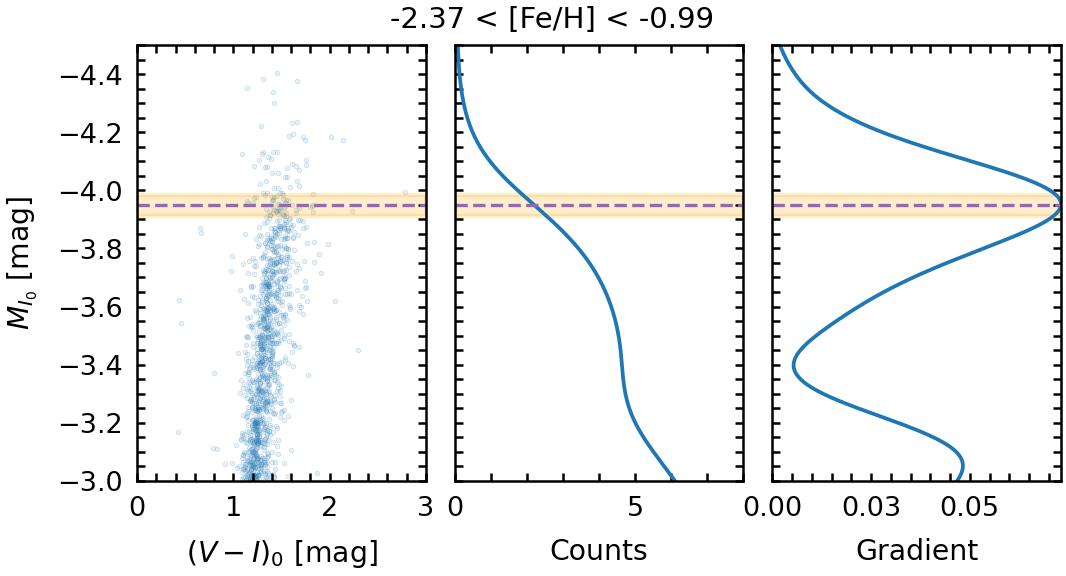}
    \vspace{-5mm}
    \caption{Color-absolute magnitude diagram, LF, and Sobel filter response for the combined GC sample based on the Stetson photometry. The purple line marks the median TRGB magnitude derived from the bootstrap analysis, and the orange contour indicates the 16–84th interquartile range.}
    \label{fig:cmd_trgb}
\end{figure}

Figure\,\ref{fig:cmd_trgb} illustrates the $I-$band CaMD combining 48 GCs and Stetson's photometry alongside the smoothed LF and Sobel filter response. Table\,\ref{tab:summary} lists our results for all three bands with central values corresponding to the median of the bootstrap samples and (asymmetric) uncertainties representing the $16-84$ percentiles. The most precise result is obtained using Stetson's GC photometry, which yields the most stars near the TRGB. The resulting $M_{I,0} =$\MIStetson\ (\MIStetsonRelativeError\ in distance) TRGB calibration is the most accurate determined in the Galaxy to date, yet agrees with previous studies performed under similar conditions.  
Specifically, our TRGB calibration agrees to within $-0.022$\,mag ($0.19\sigma$) with the maximum-likelihood analysis of field stars by \citet{2023ApJ...950...83L}, who found $M_{I,0} = -3.970^{+0.042}_{-0.024}\,(\mathrm{sys}) \pm 0.062\,(\mathrm{stat})$~mag. \citet{2021ApJ...908L...5S} previously reported $M_{I,0} = -3.97 \pm 0.04\,(\mathrm{stat}) \pm 0.10\,(\mathrm{sys})$ using GDR3 parallaxes of $\omega$~Cen. A larger difference is found with the unpublished study by \citet{2020arXiv201209701C}, where 46 GCs were combined by matching horizontal branches yielding $M_{I,0} = -4.056 \pm 0.02\,(\mathrm{stat}) \pm 0.10\,(\mathrm{sys})$. The difference of $0.108$\,mag with our determination does not exceed the systematic uncertainty reported by Cerny et al. Since this study was conducted as part of the Carnegie-Chicago Hubble Project \citep[CCHP]{2019ApJ...882...34F}, we caution that part of the difference ($\sim 0.06$\,mag) is readily explained by the weighting of Sobel filter response curves applied in the CCHP analysis. Such weighting results in brighter TRGB measurements, particularly in case of high contrast as present in GCs, cf. Appendix B in \citet{2024ApJ...963L..43A}. 

\begin{table}
\centering
\caption{TRGB calibrations derived here.}
\begin{tabular}{lccc}
\toprule
Passband & $M^{\rm TRGB}_0\,\mathrm{(mag)}$  & $N_{\text{GCs}}$ & $N^*_{[-4,-3]}$   \\
\midrule
\vspace{5pt}
$I_{\mathrm{Stetson}}$  & $-3.948^{+0.037}_{-0.034}$   & 48 & 1086 \\
\vspace{5pt}
$I_{\rm GDR3}^T$  &  $-3.949^{+0.043}_{-0.038}$ & 53 & 680 \\
$RP$  & $ -3.807^{+0.041}_{-0.035}$ & 53 & 604 \\
\bottomrule
\end{tabular}
\tablefoot{$I_{\mathrm{Stetson}}$ denotes the calibration based on ground-based photometry from the Stetson database; $I_{\rm GDR3}^T$ refers to the Cousins $I-$band calibration obtained by photometric transformations of multi-band \gaia\ data; $RP$ to native GDR3 photometry. $M^{TRGB}_{0}$ lists the median and the 16th and 84th percentiles of the bootstrap distributions derived, and the number of $N_{\text{GCs}}$ and stars in the absolute magnitude range $M_{0} \in [-4, -3]$ that passed all quality cuts (see Appendix \ref{app:data_selection}).} 
\label{tab:summary}
\end{table}

\gaia\ photometry yielded similarly accurate $I-$band (by transformation, cf. App.\,\ref{fig:photometric_transfromation_bright}) and $RP$ calibrations using 5 additional GC distances (53 instead of 48) and $\sim 40\%$ fewer stars near the TRGB due to differences in the selection criteria.  We note the near perfect agreement between the transformed and native $I-$band calibrations. We adopted the $RP$ calibration to measure distances to the LMC and SMC, relying on a single well calibrated space-based photometric system, which has become the gold standard in accurate distance scale work \citep{2025arXiv251023823H}. 

\begin{figure}
    \centering
    \includegraphics[width=.8\linewidth]{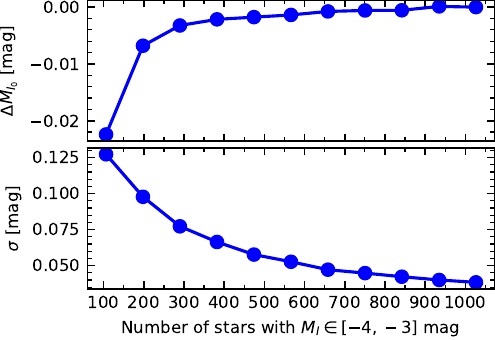}
    \caption{Stability of the TRGB measurement under random star removal. The upper panel shows the median TRGB value as a function of sample size; the median decreases as the number of stars is reduced. The lower panel shows the corresponding evolution of the TRGB uncertainty. }
    \label{fig:stability}
\end{figure}

We investigated the dependence of TRGB measurements on small-number statistics using (Stetson) $I-$band LFs and the metric $N^*_{[-4,-3]}$, the number of stars with $M_{I_{0}} \in [-4,-3]$, i.e., roughly one magnitude below the TRGB. Using simulations, \citet{1995AJ....109.1645M} had argued that $N^*_{[-4,-3]} > 100$ was required for a reliable measurement; \citet{2009ApJ...690..389M} increased this to $N^*_{[-4,-3]} > 400$. \citet{2004A&A...424..199B} pointed out this issue for GCs, and \citet[their Fig.\,10]{2022ApJ...938..101S} illustrated this effect by comparing TRGB and RR Lyrae distances to Andromeda's satellites. To measure this effect in our GC sample, we randomly removed stars from the CaMD and repeated the TRGB measurements for 11 samples with $N^*_{[-4,-3]} \in [100,1086]$. Figure~\ref{fig:stability} illustrates the deviations from the baseline and demonstrates that low statistics bias the measurement. Both the TRGB uncertainty and the bias behave asymptotically, nearing the baseline solutions for $N^*_{[-4,-3]} \gtrsim 400$, where the bias is on the order of $2\,$mmag of the baseline and the error approximately double. We stress that our approach combines information from many GCs and thus does not address the complex effect of how LFs are sampled in specific GCs \citep[as discussed in][]{2004A&A...424..199B}. 

We investigated the effect of metallicity on the TRGB calibration by grouping GCs according to [Fe/H]. Since the number of stars varies among GCs, we devised an automatic binning procedure that minimized $N^*_{[-4,-3]}$ in the largest bin. No significant trends were identified for $N_{\rm bins} \in [2,4]$. Using $N_{\rm bins}=2$, we found  $M_{I_{0,\mathrm{Stetson}}} = -3.944^{+0.056}_{-0.058}$ for [Fe/H]$=-1.91$ and  $-3.950^{+0.052}_{-0.036}$ for [Fe/H]$=-1.31$, fully compatible with each other. Finer binning resulted in ever increasing uncertainties, remaining uninformative. Our result is thus generally consistent with the expectation that old, low-metallicity populations ([Fe/H]$< -1$) exhibit no significant metallicity dependence in the $I-$band \citep{2002PASP..114..375S}. 
Metallicity calibrations from the literature suggest a brighter TRGB at lower metallicity, albeit with significant scatter between calibrations. Specifically, the widely adopted \citet{2007ApJ...661..815R} calibration predicts a $-0.017$\,mag difference based on the $(V-I)_0$ difference of $0.079$\,mag. Larger corrections of $-0.075$\,mag and $-0.056$ would be implied by the $0.6$\,dex [Fe/H] difference according to recent calibrations relying on individual GCs by \citet{2025ApJ...980..218S} and \citet{2026RAA....26c5018Y}, respectively. However, such calibrations are dominated by few GCs at high metallicity, differ in their mathematical form (exponential or polynomials) and can be affected by the aforementioned stochastic issues. 

While a TRGB metallicity dependence thus cannot be ruled out, we conclude that a) this effect cannot be precisely constrained using this collection of GCs due to the generally low metallicity range and that b) our calibration benefits from insensitivity to stochastic sampling effects affecting analyses of individual GCs.

\section{Magellanic Cloud Distances to $2\%$} \label{sec:distance_lmc}
The distances to the Magellanic Clouds play crucial roles for the calibration of various standard candles and the extragalactic distance scale. The most commonly adopted distances to both galaxies, $\mu_{\mathrm{LMC}} = 18.477 \pm 0.004\, (\mathrm{stat}) \pm 0.026\,(\mathrm{sys})$~mag ($1.1\%$ in distance) in the LMC and $\mu_{\mathrm{SMC}} = 18.977 \pm 0.016\,(\mathrm{stat}) \pm 0.028\,(\mathrm{sys})$~mag ($1.5\%$ in distance) in the SMC, have been determined geometrically using DEB systems composed of helium-burning giants \citep{2019Natur.567..200P,2020ApJ...904...13G}. These distances have also been considered in the Local Distance Network to measure the Hubble constant \citep{2025arXiv251023823H}. Measuring both distances independently of the DEBs and with competitive accuracy thus provides crucial cross-checks. 

Our \MRpRelativeError\ TRGB calibration using \gaia's $RP$ observations provides such an opportunity, benefiting from a single well-calibrated space-based photometric system, an absolute scale provided by trigonometric parallaxes, and even taking into account the correlated nature of the GC distances. We thus determined  $\mu_{\mathrm{LMC}} =$ \muLMCRp\ (\relMuLMCRp\ in distance) and $\mu_{\mathrm{SMC}}=$\muSMCRp\ (\relMuSMCRp\ in distance) using our value for $M_{RP,0}$ and the apparent magnitudes for the small amplitude red giants (SARGs) samples from Tab. 3 in \citet{2024ApJ...974..181K}. 
We adopted SARG-based LMC TRGB measurements because a) all stars near the TRGB are SARGs \citep{2024ApJ...963L..43A,2024ApJ...974..181K}, b) the variability selection filters out contaminants similar to the astrometric membership in GCs, and c) the SARG LFs in the LMC and SMC are inherently less sensitive to smoothing bias than the Allstars samples, similar to the combined LF constructed from the GCs (cf. App.\,\ref{sec:smoothing} and Fig.\,\ref{fig:smoothing}). Nevertheless, we stress that differences between the SARG and Allstars TRGB magnitudes in the LMC and SMC are negligible: $5\,$mmag and $1\,$mmag, respectively.
For the LMC, our distance agrees to within $<0.7\sigma$ or $\Delta \mu_{\mathrm{DEB-TRGB,LMC}} = 0.030^{+0.050}_{-0.045}$~mag with the DEB result. Analogously for the SMC, we find agreement within $\sim 1.3\sigma$  or $\Delta \mu_{\mathrm{DEB-TRGB,SMC}} = 0.079^{+0.063}_{-0.059}$~mag with the DEB distance. 
The slightly worse agreement arises because the SMC-LMC TRGB magnitude difference of $0.450$\,mag ($0.446$\,mag in $I-$band) is slightly smaller than the $0.500$\,mag difference among the DEB distances \citep[cf. Sect.\,3.2 in][]{2024ApJ...974..181K}.

Applying color-based metallicity corrections from \citet{2007ApJ...661..815R} would tend to slightly increase our distances. While these calibrations were determined in the $I-$band, we apply them here to \gaia\ $RP$ measurements; this is warranted by the similar effective wavelengths, which differ by merely $9$\,nm \citep{2010A&A...523A..48J}.
Using the mean dereddened color of the GC sample near the TRGB, $(V-I)_{0,\mathrm{GC}} = 1.49$, and the values for SARGs in the LMC and SMC yields color differences of $0.31$\,mag and $0.09$\,mag, respectively. This would yield $\Delta \mu_{\mathrm{DEB-TRGB,LMC}}^{\rm Zcorr} = \Delta \mu_{\mathrm{DEB-TRGB}} - 0.217\left[(V-I)_{0,\mathrm{LMC}}-(V-I)_{0,\mathrm{GC}}\right]
= -0.037\, \mathrm{mag}$ and $\Delta \mu_{\mathrm{DEB-TRGB,SMC}}^{\rm Zcorr} = 0.059$\,mag. Both of our TRGB distances are consistent with the DEBs after applying such corrections, although the changes are hardly significant. 

The \citet{2025arXiv251023823H} recently presented the Local Distance Network (LDN), a new framework for determining the Hubble constant by combining a large collection of high-quality distance measurements in a statistically rigorous manner. Specifically, the LDN included the DEB distances to the LMC in the baseline solution and that to the SMC in variant V07 (baseline+SMC). Since our TRGB distances are fully independent of the DEB distances, we replaced the two DEB distances by the weighted averages of both methods, $\langle \mu_{LMC} \rangle = 18.468 \pm 0.022$\,mag ($1.02\%$ in distance) and  $\langle \mu_{SMC} \rangle = 18.955 \pm 0.027$\,mag ($1.25\%$ in distance), assuming Gaussian errors (mean of the asymmetric uncertainties) to match the design of the LDN. Using the LDN code soon to be released, we determined $H_0 = 73.52\pm 0.80\, \mathrm{km\,s^{-1}\,Mpc^{-1}}$ and $73.29\pm 0.78 \,\mathrm{km\,s^{-1}\,Mpc^{-1}}$ for the modified baseline and V07, respectively. As expected, no significant change in $H_0$ is found, and uncertainties are only marginally improved. Nevertheless, including this information increases the redundancy of the LDN and demonstrates how new and improved constraints may included in the LDN in the future.

\section{Conclusions}\label{sec:conclusions}
We have calibrated the TRGB using $48$ and $53$ GCs in $I-$band and \gaia's $RP$ system, achieving the greatest accuracy to date for the Milky Way TRGB calibration. The absolute scale is provided by a purely \gaia-based joint solution to $802$ RRab, $345$ RRc, and $21$ type-II Cepheids residing in Galactic GCs anchored to average GC parallaxes \citep{2025arXiv250916331L}. Our $I-$band calibration based on ground-based data from the Stetson database yields \MIStetson\ (\MIStetsonRelativeError\ in distance) in full agreement with previous studies, albeit with much reduced uncertainties. Stated uncertainties include statistical and systematic effects, including the correlated nature of the underlying GC distances. Additional investigations of the stability of the measurements against low statistics, of a possible dependence on the smoothing parameter, of reddening and metallicity effects were presented. In particular, we conclude that our set of Galactic GCs does not exhibit a significant metallicity dependence of the TRGB, limited by either a restricted [Fe/H] range or by TRGB measurement bias and increased uncertainties resulting from stochastic effects. Conversely, we showed that color-based metallicity corrections from the literature would not significantly affect our results.

Our \gaia\ $RP$ calibration yields \MRp\ (\MRpRelativeError\ in distance) and offers the opportunity of measuring TRGB distances to the LMC and SMC with a single, well-calibrated space-based photometric system and based on an absolute scale anchored to geometric distances. Using consistently measured TRGB apparent magnitudes from \citet{2024ApJ...974..181K}, we obtained \muLMCRp\ (\relMuLMCRp\ in distance) and \muSMCRp\ (\relMuSMCRp\ in distance) in the LMC and SMC, respectively. These distances are entirely independent of, yet fully compatible with, the high-accuracy geometric DEB distances  \citep{2019Natur.567..200P,2020ApJ...904...13G}. Incorporating the weighted average of the new TRGB and existing DEB distances in the LDN as an example of its use yields a very slightly improved uncertainty on $H_0=73.52\pm0.80\, \mathrm{km\,s^{-1}\,Mpc^{-1}}$ for the updated LDN baseline. 

Several significant improvements to this calibration will be enabled by the upcoming fourth \gaia\ data release (GDR4), currently scheduled for December 2026. GDR4 will feature improved astrometric solutions, photometric calibrations, and variable star classifications, improving all aspects of this work.

Additionally, GDR4 will enable studies of SARGs in GCs in analogy with the Magellanic Clouds \citep{2024ApJ...963L..43A}. Crowding limited both the number of GCs and stars therein that could be used in the present work. Improving photometric measurements in crowded regions and at the bright end of the GCs would thus be particularly valuable. Further improvements are likely thanks to homogeneous and detailed abundance measurements in GCs and the Magellanic Clouds based on the 4-metre Multi-Object Spectroscopic Telescope (4MOST) \citep{4MOST,4MOSTS9,4MOSTclusters}, whereas differential reddening maps \citep{2024A&A...686A.283P} may further sharpen CaMDs.

\begin{acknowledgements}
MC and RIA were funded by a Swiss National Science Foundation Eccellenza Professorial Fellowship (award PCEFP2\_194638). This research has received support from the European Research Council (ERC) under the European Union's Horizon 2020 research and innovation programme (Grant Agreement No. 947660).  

This work has made use of data from the European Space Agency (ESA) mission {\it Gaia} (\url{https://www.cosmos.esa.int/gaia}), processed by the {\it Gaia} Data Processing and Analysis Consortium (DPAC,
\url{https://www.cosmos.esa.int/web/gaia/dpac/consortium}). Funding for the DPAC has been provided by national institutions, in particular the institutions participating in the {\it Gaia} Multilateral Agreement. 

\end{acknowledgements}

\bibliographystyle{aa}
\bibliography{refs}

\begin{appendix}

\section{Photometry}\label{sec:photometry}
Peter Stetson's website notes that the photometry of the brightest objects in each field may be affected by saturation and cautioned that such measurements should not be used without external verification of their quality. This appendix addresses that concern and shows that saturation does not represent an issue for our analysis. This point is particularly relevant here, as the brightest stars are precisely those used to measure the TRGB. 

\begin{figure}[ht]
    \centering
\includegraphics[scale = 1]{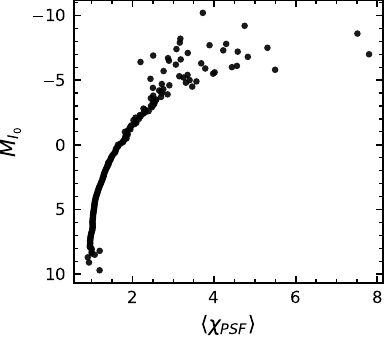}
\caption{Mean reduced $\chi_{\mathrm{PSF}}$ as a function of absolute magnitude in the $I$ band, computed in bins of 0.1 mag.}
    \label{fig:chi_square_psf}
\end{figure}

We first estimated the absolute magnitude of each star using the parameters described in Sect.~\ref{sec:data} and grouped stars from all clusters into bins of 0.1~mag. The results are shown in Fig.~\ref{fig:chi_square_psf}, which illustrates how the reduced chi-square ($\chi_{\mathrm{PSF}}$) of the point-spread-function (PSF) fitting varies as a function of absolute magnitude in the $I$ band. As shown, $\chi_{\mathrm{PSF}}$ reaches its maximum for the brightest stars and has a value close to 3 near the TRGB. It is worth noting that this behavior is limited to the $I$-band photometry from Stetson and has no impact on the Gaia $RP$ band.

To assess the accuracy of these measurements, we employed two independent checks. The first uses Gaia synthetic photometry (Sect. \ref{sec:gaia}), and the second relies on photometric transformations derived from faint stars (Sect. \ref{sec:transformations}). Both tests indicate that the photometry is reliable.

\subsection{Gaia synthetic photometry}\label{sec:gaia}
To evaluate the quality of the Stetson photometry at the bright end, we computed Gaia $I$-band synthetic photometry \citep{2023A&A...674A..33G} in the Johnson–Kron–Cousins system (“JKC\_Std”) for all GCs. The synthetic photometry was standardized using the \citet{2019MNRAS.485.3042S} photometric standards, however, since these standards are fainter than the stars used to measure the TRGB, there is no a priori guarantee that the two photometric sets agree for bright sources.

\begin{figure}[ht]
    \centering
    \includegraphics[scale = 0.85]{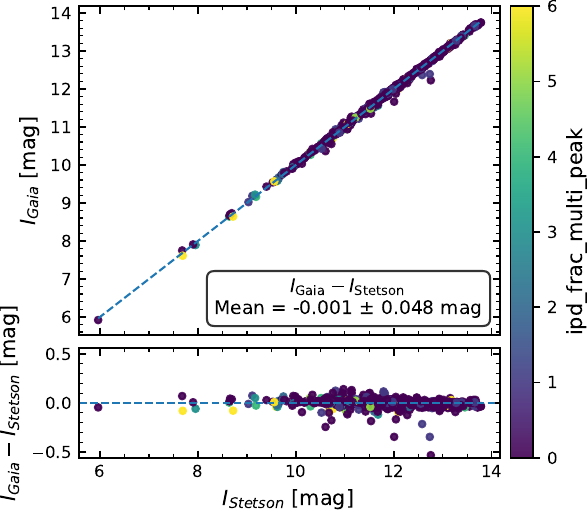}
    \caption{Comparison between Gaia synthetic photometry and Stetson photometry. No systematic bias is observed for stars brighter than the photometric standards, indicating that the Stetson photometry can be reliably used to measure the TRGB.}
    \label{fig:synthetic}
\end{figure}

We compared the sources for which both measurements are available. For this comparison, we use only sources brighter than the top 5\% of the photometric standards in each field. We removed sources with $\lvert \mathrm{sharp} \rvert > 1$ or $\texttt{ipd\_frac\_multi\_peak} > 10$, clusters with high differential reddening, and clusters whose color–magnitude diagrams appear contaminated by background stars.  The results are shown in Fig. \ref{fig:synthetic}. The mean difference of $I_{\mathrm{Gaia}} - I_{\mathrm{Stetson}}$ is $-0.001 \pm 0.048$ mag, where the uncertainty represents the standard deviation. Thus, we do not find evidence for a systematic bias in the Stetson photometry for bright sources. We decided not to use the synthetic photometry for the TRGB measurement, because only $\sim 12\%$ of the stars in the Stetson sample have such measurements available.

\subsection{Photometric transformations}\label{sec:transformations}
A second check of the Stetson photometry for bright stars is obtained from photometric transformations calibrated using faint stars. To derive the $I$-band transformation, we use Gaia $BP$ and $RP$ photometry and exclude all stars brighter than the top 5\% of the photometric standards in each field. We further apply the photometric quality cuts described in Appendix~\ref{app:data_selection}. The transformation is modeled as $I_{\mathrm{pred}} \equiv RP + f(C)$, where $f(C)$ is a third-order polynomial of the form $f(C) = \sum_{k=0}^{3} a_{k} C^{k}$, with $C = BP - RP$. The propagated photometric uncertainty for each star is computed via standard error propagation as

\begin{equation}
\sigma^{2} = \left(1 - f'(C)\right)^{2}\sigma_{RP}^{2} + \left(f'(C)\right)^{2}\sigma_{BP}^{2},    
\end{equation}
where $f'(C)$ denotes the derivative of the polynomial with respect to $C$.  The coefficients of the transformation appear in Table \ref{tab:coefficients}. 

\begin{table}[ht]
\centering
\caption{Coefficients of the third–order polynomial fit.}
\begin{tabular}{lcc}
\hline\hline
Coefficient & mean  \\
\hline
$a_0$ & -0.003281  \\
$a_1$ & -0.105225  \\
$a_2$ & 0.008891  \\
$a_3$ & 0.003471\\
\hline
\end{tabular}
\label{tab:coefficients}
\end{table}

\begin{figure}[ht]
    \centering
    \includegraphics[scale=0.85]{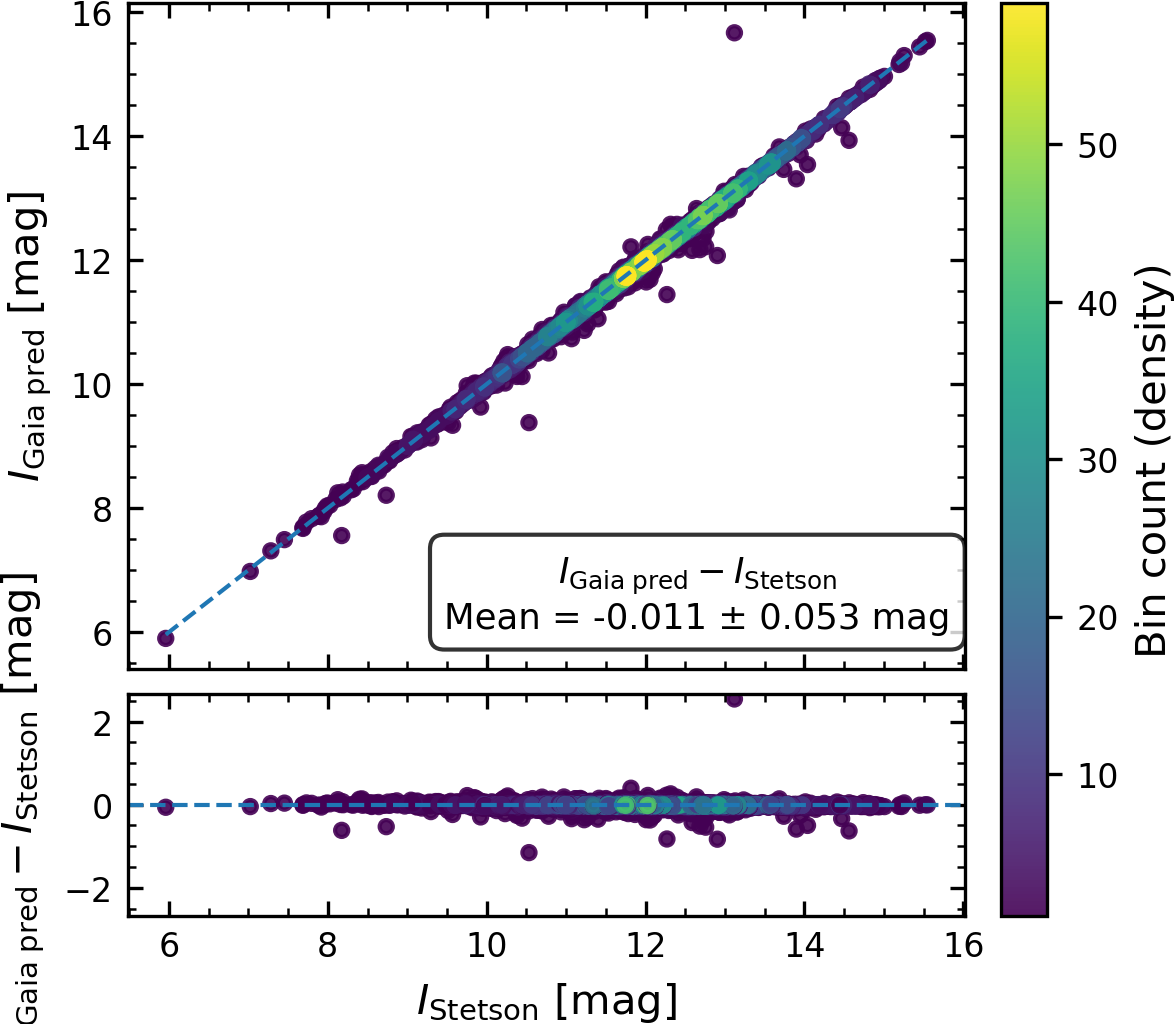}
    \caption{Comparison between the Stetson photometry of stars brighter than the top 5\% of the standards and the photometry obtained by applying transformations calibrated using fainter stars.}
\label{fig:photometric_transfromation_bright}
\end{figure}

The comparison between the photometry for sources brighter than the top 5\% of the photometric standards (and therefore not used in the calibration of the transformations) is shown in Fig. \ref{fig:photometric_transfromation_bright}. The difference between the two measurements is statistically consistent with zero, and therefore there is no evidence of a photometric bias.

\section{Data selection criteria}\label{app:data_selection}
This appendix summarizes the full set of quality cuts applied to the photometric data, Gaia astrometry, and cluster-level parameters. These filters were designed to remove objects with poor photometry, unreliable Gaia measurements, or uncertain cluster properties.

We began with 93 clusters with measured distances \citep{2025arXiv250916331L}. The reddening cut, $E(B-V) < 1$, removed 5 clusters. The metallicity cut, $[\mathrm{Fe/H}] < -0.9$, excluded 7 additional clusters. Requiring more than 100 stars per cluster removed 9 more, and the condition $n_{\mathrm{stars,fit}} > 2$ excluded 14 clusters. In addition, 5 clusters were removed because their color--magnitude diagrams exhibit strong differential reddening, or are significantly affected by contamination from foreground or background stars (NGC~6171, NGC~6266, NGC~6401, NGC~6626, and NGC~6715).  After applying all criteria, the final sample contains 53 clusters, of which 48 have photometry in the $I$ band. The adopted quality cuts are summarized in Table~\ref{tab:selection_criteria}, and the clusters used for the calibration in each band are listed in Table~\ref{tab:clusters}.

\begin{table*}[ht]
\centering
\setlength{\tabcolsep}{6pt} 
\caption{Quality cuts.}
\begin{tabular}{ccc}
\hline\hline
\textbf{Stetson photometry} & \textbf{Gaia photometry} & \textbf{Cluster parameters} \\
\hline
$\chi_{\mathrm{PSF}} < 4$ & $\texttt{ipd\_frac\_multi\_peak} < 10$ & $E(B-V) < 1$ \\
$\lvert \mathrm{sharp} \rvert < 1$ & $C^{*}/\sigma_{C^{*}} < 3$ & $\mathrm{[Fe/H]} < -0.9$ \\
$I < 18$ &  & $N_{\mathrm{stars,\,cl}} > 100$\\
 &  & $N_{\mathrm{stars,\,distance}} > 2$ \\
 &  & $\mathrm{Membership\ probability} > 0.5$ \\
\hline
\end{tabular}
\label{tab:selection_criteria}
\tablefoot{The parameters $\chi_{\mathrm{PSF}}$ and $\mathrm{sharp}$ were defined in \citet{2019MNRAS.485.3042S}. The definition of \texttt{ipd\_frac\_multi\_peak} is given in the Gaia archive documentation, and that of the $C^{*}$ parameter was given in \citet{2021A&A...649A...3R}. The $E(B-V)$ and [Fe/H] values were taken from \citet{2010arXiv1012.3224H}. The number of cluster members ($N_{\mathrm{stars,,cl}}$) were taken from \citet{2024A&A...684A.173C}. The number of stars used in the Leavitt law fit per cluster ($N_{\mathrm{stars,,distance}}$) was taken from \citet{2025arXiv250916331L}, while the membership probabilities were taken from \citet{2021MNRAS.505.5978V}.}

\end{table*}

\begin{table}[ht]
\centering
\caption{Clusters used for the TRGB calibration in each band. The full list is available in electronic form at the CDS.}
\label{tab:clusters}
\begin{tabular}{lccc}
\hline
Cluster & $I_{\mathrm{Stetson}}$  & $I^{\mathrm{T}}_{\mathrm{GDR3}} $  & $RP$ \\
\hline
NGC~6121 & 1 & 1 & 1 \\
NGC~6656 & 1 & 1 & 1 \\
NGC~3201 & 1 & 1 & 1 \\
$\cdots$ & $\cdots$ & $\cdots$ & $\cdots$ \\
\hline
\end{tabular}
\tablefoot{A value of 1 indicates that the cluster is included in the calibration in the corresponding band, while 0 indicates it is not.}
\end{table}

\section{Sensitivity to smoothing}\label{sec:smoothing}
\begin{figure}[ht]
    \centering
\includegraphics[width=1\linewidth]{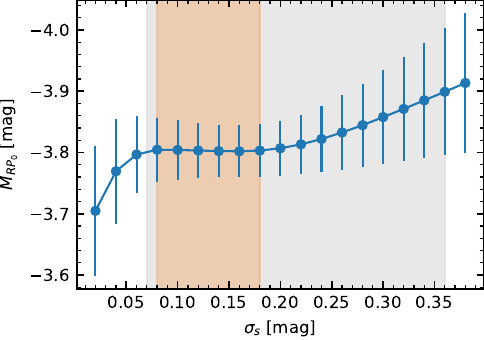}
    \caption{Variation of the mean and standard deviation of the TRGB magnitude as a function of the GLOESS smoothing parameter. The grey vertical lines indicate the range over which the absolute magnitude of the TRGB is insensitive to the smoothing value, according to \citet{2024ApJ...963L..43A} for their “SARGs” sample. The yellow shaded region was determined for the GC sample using the criteria $\left| \frac{d M_{RP_{0}}}{d \sigma_s} \right| \le 0.10
$ and covers the range $\sigma_{s} \in [0.08, 0.18]$ mag. }
    \label{fig:smoothing}
\end{figure}

Previous studies \citep{2024ApJ...963L..43A,2025AcA....75....1U} have shown that the choice of the GLOESS smoothing parameter can introduce systematic biases in the TRGB determination. To quantify the impact of this effect, we repeated the TRGB measurement in the $I_{\mathrm{Stetson}}$ band over a range of smoothing values, $\sigma_{s} \in [0.10, 0.38]$~mag, in steps of 0.02~mag. The results are shown in Fig.~\ref{fig:smoothing}. Relative to the reference value adopted in Sect.~\ref{sec:results}, $\sigma_{s} = 0.125$, adopting $\sigma_{s} = 0.15$ shifts the mean TRGB to fainter magnitudes by 0.004~mag, whereas $\sigma_{s} = 0.10$ shifts it to brighter magnitudes by 0.005~mag. These offsets provide an estimate of the systematic uncertainty associated with the choice of the smoothing parameter.  

\section{Uncertainty propagation and convergence}\label{sec:convergence}
The distances reported by \citet{2025arXiv250916331L} are mutually correlated. To propagate these correlations through the TRGB analysis, we used the MCMC samples presented in their Sect. 4.1.1, which represent the joint posterior distribution of the cluster distances. After discarding the burn-in phase and thinning the chains, we retained a set of $P$ approximately independent MCMC realizations. Each realization corresponds to one instance of the full distance vector and therefore preserves the distance correlations. For each MCMC realization, we performed $Q$ bootstrap resamplings of the stars in each cluster to account for sampling variance. This procedure yielded a total of $P \times Q$ realizations. For each realization, we measured the TRGB absolute magnitude. The adopted TRGB value corresponds to the median of the resulting distribution, and the uncertainties are defined by the 16th and 84th percentiles.

To assess the convergence of the sampling procedure, we analyzed the sequence of TRGB measurements. We compute the cumulative mean of this sequence, $\mu$, and define the convergence point $k_{\mathrm{conv}}$ as the first iteration at which the mean becomes stable. Specifically, we require that the last $M$ values of $\mu$ remain within $\epsilon$ of $\mu_k$, i.e.,
\begin{equation}
k_{\mathrm{conv}} =
\min\left\{
k \ge M \,:\,
\max_{j \in \{k-M+1,\ldots,k\}}
\left|\mu_j - \mu_k\right| < \epsilon
\right\}.
\end{equation}
We first quantified how many MCMC samples are required to reach convergence. We found that $N_{\mathrm{MCMC}} = 1,500$ samples are sufficient to achieve convergence at the $\epsilon = 0.001$~mag level, adopting a stability window of $M = 400$ steps, and it is reached at $k_{\mathrm{conv}} \approx 666$. 

\begin{figure}[H]
\centering        \includegraphics[width=1\linewidth]{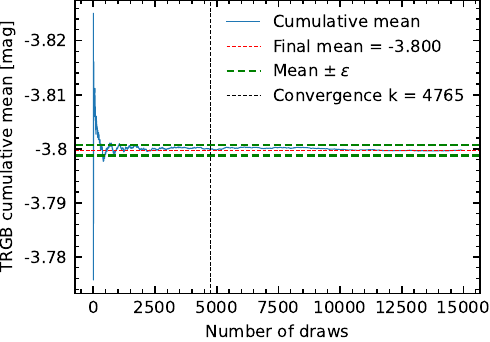}
    \caption{Convergence of the TRGB estimates in the RP-band. The blue curve shows the cumulative mean TRGB absolute magnitude as a function of the number of sample draws. The red dashed line marks the final mean of the full sample. The green dotted lines indicate the final mean $\pm \epsilon$, with $\epsilon = 0.001$ mag, and the black vertical line marks the convergence index $k_{\mathrm{conv}}$. As discussed in the text, the sample draws were randomly reordered; therefore, the exact shape of the curve may vary between permutations.}
    \label{fig:convergence_plot}
\end{figure}

Having established convergence with the MCMC draws, we next evaluated the full sampling procedure, including the bootstrap resamplings. After generating 10 bootstrap resamples for each MCMC realization and re-measuring the TRGB, we assessed convergence of the combined sequence of measurements. Since the realizations are naturally processed in order (i.e., the procedure is first applied to the first MCMC draw, then to the second, and so on), applying the convergence criterion to the ordered sequence could lead to convergence before incorporating information from the final realizations. To avoid this, we randomly shuffled the combined set of bootstrap measurements and evaluated convergence on the randomized sequence. We find that, on average, the combined sequence reached convergence at $k_{\mathrm{conv}} \approx 5618$ at the $\epsilon = 0.001$~mag level, using $M = 4{,}000$. An example of a full run in the RP-band is shown in Fig.~\ref{fig:convergence_plot}.

\end{appendix}

\end{document}